\documentclass[superscriptaddress,a4paper,english,notitlepage,12pt,tightenlines,nofootinbib]{revtex4-2}
\usepackage[]{graphicx}
\usepackage{amsmath,amssymb}
\usepackage{newtxtext}
\usepackage{newtxmath}
\usepackage{verbatim}
\usepackage[utf8]{inputenc}
\usepackage[x11names]{xcolor}
\usepackage[unicode,colorlinks,citecolor=Blue3,linkcolor=Blue3,bookmarks=true]{hyperref}

\newcommand{\ie}{{\it i.e.\,}}
\newcommand{\eg}{{\it e.g.\,}}
\newcommand{\mean}[1]{\langle{#1}\rangle}
\newcommand{\partd}[2]{\dfrac{\partial#1}{\partial#2}}
\newcommand{\Det}{\mathop{\rm Det}\nolimits}
\newcommand{\smatrix}[4]{\begin{pmatrix}#1 & #2 \\ #3 & #4\end{pmatrix}}
\newcommand{\Tr}{\mathop{\rm Tr}\nolimits}
\newcommand{\bra}[1]{\langle{#1}|}
\newcommand{\ket}[1]{|{#1}\rangle}

\begin{document}


\title{Sensitivity of quantum-enhanced interferometers}

\author{D.~Salykina}

\affiliation{Lomonosov Moscow State University, 119991 Moscow, Russia}
\affiliation{Russian Quantum Center, Skolkovo 143025, Russia}%

\author{F.~Ya.~Khalili}

\email{farit.khalili@gmail.com}
\affiliation{Russian Quantum Center, Skolkovo IC, Bolshoy Bulvar 30, bld.\ 1, Moscow, 121205, Russia}
\affiliation{NTI Center for Quantum Communications, National University of Science and Technology MISiS, Leninsky prospekt 4, Moscow 119049, Russia}


\begin{abstract}
  We consider various configuration of quantum-enhanced interferometers, both linear (SU(2)) and non-linear (SU(1,1)) ones, as well as hybrid SU(2)/SU(1,1) schemes.
  Using the unified modular approach, based on the Quantum Cram\`er-Rao bound, we show that in all practical cases, their sensitivity is limited by the same equations  \eqref{conc_phi} or \eqref{conc_phi_pm} which first appeared in the pioneering work by C.\,Caves \cite{Caves1981}.
\end{abstract}

\maketitle


\section{Introduction}

\begin{figure}
  \includegraphics{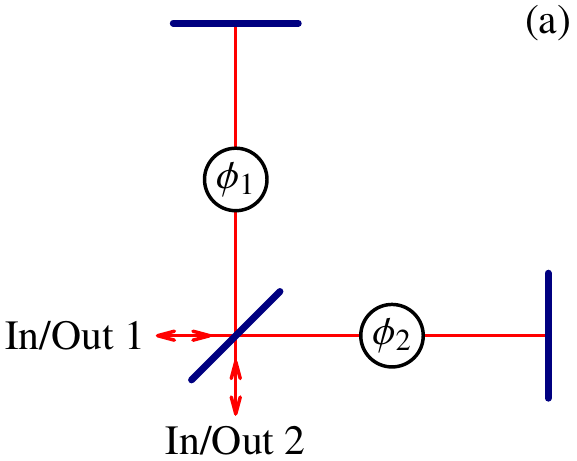}\quad
  \includegraphics{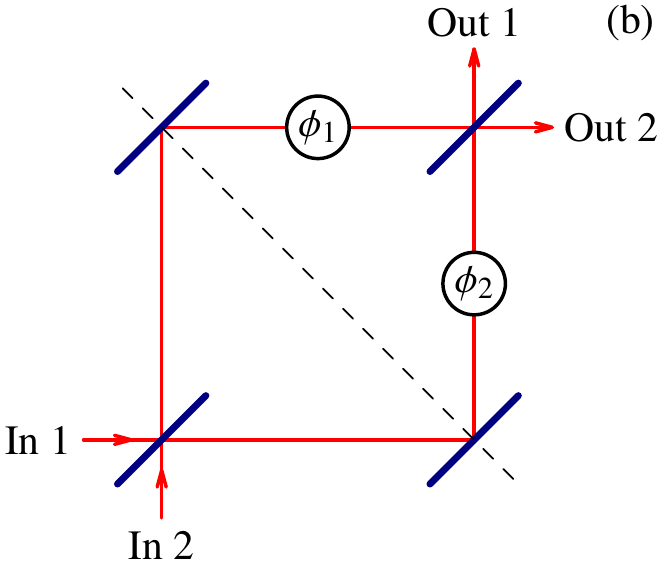}
  \caption{Michelson (a) and Mach-Zehnder (b) interferometers.}\label{fig:SU2}
\end{figure}

Optical interferometers are an indispensable tool for many scientific and industrial applications. In particular, two famous interrelated examples worth mentioning here. In 1987, the Michelson interferometer (see Fig.\,\ref{fig:SU2}(a)) helped to disprove the ether theory \cite{Michelson1887}, paving the way to A.\,Einstein's special theory of relativity. After more than one century, the first direct observation of the gravitational waves \cite{PRL_116_061102_2016} by the pair of LIGO interferometers \cite{LIGOsite} provided one of the most convincing proofs of A.\,Einstein's general theory of relativity.

Sensitivity of the best modern interferometers is very high. Probably, they are the most sensitive measurement devices available now. For example, the contemporary laser interferometric gravitational-wave detectors, like LIGO and VIRGO \cite{VIRGOsite}, can measure relative elongations of their 3-4 km arms with the precision  exceeding $\sim10^{-23}\,{\rm Hz}^{-1/2}$ \cite{Tse_PRL_123_231107_2019_short}. The major factor which limits their sensitivity is the quantum noise of the probing light. In the most basic case of a coherent quantum state of light, the corresponding limit is known as the shot-noise one (SNL):
\begin{equation}\label{SNL}
  \Delta\phi_{\rm SNL} = \frac{1}{2\sqrt{N}} \,,
\end{equation}
where $N$ is the mean number of photons interacting with the phase shifting object(s) and $\phi$ is the measured phase.

This limit suggests the straightforward way of improving the phase sensitivity, namely the increase of $N$. However, there are important cases where this brute-force approach can not be used. In particular, in the laser gravitational-wave detectors, the circulating optical power reaches hundreds of kilowatts and is limited by various undesired nonlinear effects, like thermal distortions of the mirrors shape or optomechanical parametric excitation \cite{CQG_32_7_074001_2015, Richardson_2201_09482}. Another example is the biological measurements \cite{Taylor_PR_615_1_2016}, where the probing light intensity could be limited due to the fragility of the samples.

At the same time, it is known that the limit \eqref{SNL} can be overcome using more sophisticated ``non-classical'' quantum states. The simplest and the only practical, accounting for the contemporary technological limitations, class of such states consists of the Gaussian quadrature-squeezed states, which differ from the coherent ones by decreased by $e^{-r}$ (where $r$ is the logarithmic squeeze factor) uncertainty of one of two its quadratures and proportionally increased uncertainty of another one \cite{Yuen_PRA_13_2226_1976}. It was shown in the pioneering work by C.\,Caves \cite{Caves1981}, that in the realistic case of moderate squeezing, $e^{2r}\ll N$, the phase sensitivity can be improved by the factor $e^{-r}$:
\begin{equation}\label{SQZ}
  \Delta\phi_{\rm SQZ} \approx \frac{e^{-r}}{2\sqrt{N}} \,.
\end{equation}
In the hypothetical opposite case of very strong squeezing, $e^{2r}\sim N$, the phase sensitivity could approach the so-called Heisenberg limit (HL) \cite{Holland_PRL_71_1355_1993, Lane_PRA_47_1667_1993, Sanders_JMO_44_1309_1997, Demkowicz-Dobrzanski_NComm_3_1063_2012, Pezze_PRA_91_032103_2015}, which, in the asymptotic case of $N\gg1$, can be presented as follows:
\begin{equation}\label{HL}
  \Delta\phi_{\rm HL} \sim \frac{1}{N} \,.
\end{equation}

Sensitivity exceeding the SNL \eqref{SNL} can be achieved also using exotic non-Gaussian quantum states, like Fock states \cite{Pezze_PRL_110_163604_2013}, twin Fock states \cite{Holland_PRL_71_1355_1993, Campos_PRA_68_023810_2003}, NOON states \cite{Lee_JMO_49_2325_2002, Berry_PRA_80_052114_2009} or multi-mode Fock states \cite{Perarnau-Llobet_QST_5_025003_2020}; see also the review \cite{Demkowicz_PIO_60_345_2015}. The sensitivity saturating the HL using a non-Gaussian quantum state was experimentally demonstrated in Ref.\,\cite{Daryanoosh_NComm_9_4606_2018}, but for very modest photons number $N=3$.

However, generation of these states with $N\gg1$ is problematic with the current state-of-the-art technologies. Moreover, it was shown in Refs.\,\cite{Lang_PRL_111_173601_2013, Lang_PRA_90_025802_2014}, that in the important case of a symmetric interferometer with 50\%/50\% input beam splitter(s), the optimal sensitivity could be provided by the Gaussian squeezed states.

The first proof-of-principle experiments with the squeezed-light-enhanced interferometers  were done in 1987 \cite{Xiao_PRL_59_278_1987, Grangier_PRL_59_2153_1987}. In the recent ``table-top'' scale experiments, squeezing up to 15\,db was demonstrated \cite{Eberle_PRL_104_251102_2010, 20a1FrAgKhCh, Zahnder_thetis2021, Vahlbruch_PRL_117_110801_2016, Mehmet_CQG_36_015014_2019, 20a1FrAgKhCh, Darsow-Fromm_OL_46_5850_2021, Heinze_PRL_129_031101_2022}. In 2011, the squeezing was implemented in the relatively small gravitational-wave detector GEO-600 \cite{Nature_2011}, and in 2019 --- in the larger and more sensitive detectors Advanced LIGO \cite{Tse_PRL_123_231107_2019_short} and Advanced VIRGO \cite{Acernese_PRL_123_231108_2019_short}, providing the sensitivity gain of about 3\,dB. The gain up to 10\,dB is planned for the future so-called third generation detectors, like Einstein Telescope \cite{Punturo_CQG_27_194002_2010} and Cosmic Explorer  \cite{Reitze_1907_04833}.

The main factor which limits the gain promised by the non-classical light is the optical losses. The non-Gaussian states, like the Fock or NOON ones, are most readily destroyed by them \cite{Zurek_RMP_75_715_2003}. The Gaussian squeezed states are more robust.  But nevertheless, in their case the losses also very significantly limit the sensitivity. As it was shown in Ref.\,\cite{Demkowicz_PRA_88_041802_2013}, in presence of losses, the effective squeezing is limited by the following relation:
\begin{equation}\label{r_loss}
  e^{r_{\rm loss}} = \sqrt{\frac{\eta}{1-\eta}} \,,
\end{equation}
where $\eta$ is the overall quantum efficiency of the interferometer.

The standard topologies of the Michelson and Mach-Zehnder interferometers used in the high-precision phase measurements  are shown in Figs.\,\ref{fig:SU2}(a) and \ref{fig:SU2}(b), respectively. 
Evidently, the scheme of Fig.\ref{fig:SU2}(a) can be reduced to the one of \ref{fig:SU2}(b) by folding along the dashed line. Therefore, the Michelson interferometer is equivalent to the particular case of the Mach-Zehnder with the identical input and output beamsplitters. In Ref.\,\cite{Yurke_PRA_33_4033_1986}, the term ``SU(2) interferometer'' was coined for them because the operations of passive beam splitters and phase shifters can be visualized as rotations of the Stokes vectors described by the group SU(2).

In the standard operating regime of the SU(2) interferometers, one of the input ports (the {\it bright} one) is excited by the laser light. Within the idealized theoretical treatment, this corresponds to injection of a coherent quantum state into this port. The second port either remains dark, or the squeezed vacuum is fed into it, as it was proposed in Ref.\,\cite{Caves1981}. Evidently, other combinations with coherent, squeezed vacuum, or squeezed coherent quantum states being injected into one or both input ports are possible, see \eg Refs.\,\cite{Lang_PRA_90_025802_2014, Ataman_PRA_98_043856_2018, Ataman_PRA_100_063821_2019, Ataman_PRA_102_013704_2020} and the review \cite{Anderson_ch35_2019}.

\begin{figure}
  \includegraphics{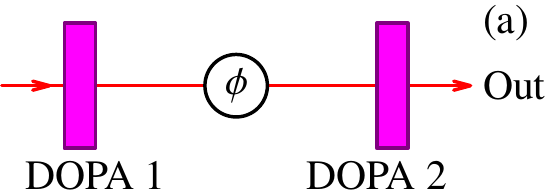}\quad
  \includegraphics{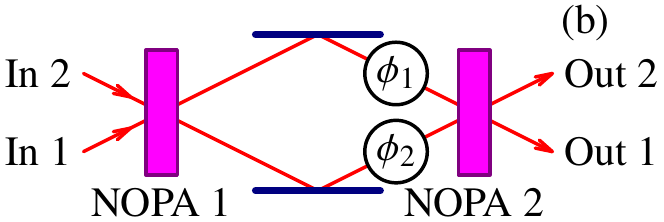}
  \caption{Single-arm (a) and two-arm (b) SU(1,1) interferometers. DOPA~1,2: degenerate optical parametric amplifiers; NOPA: nondegenerate optical parametric amplifiers.}\label{fig:SU11}
\end{figure}

In 1986, a novel scheme of the so-called SU(1,1) interferometer was proposed by Yurke et al.~\cite{Yurke_PRA_33_4033_1986}, see Fig.\,\ref{fig:SU11}. The key elements here are two optical parametric amplifiers (OPA), which replace the beamsplitters used in the SU(2) interferometer. The OPAs could be either degenerate (DOPA), that is with the signal and idler beems being identical, or non-degenerate (NOPA), with the physically separated signal and idler beams.  The name ``SU(1,1) interferometer'' is due to the fact that the action of parametric amplifiers can be described in therms of the Lorentz transformation.

At first sight, the non-degenerate version shown in Fig.\,\ref{fig:SU11}(b) looks very similar to the Mach-Zehnder interferometer (Fig.\,\ref{fig:SU2}), with the input and output squeezers playing the role of the respective beamsplitters. However, this visual similarity is misleading. Opposite to the SU(2) interferometers the non-degenerate SU(1,1) one is sensitive only to the sum phase shift $\phi_1+\phi_2$ in the arms relative to the parametric pump phase, see Sec.\,\ref{sec:SU11_meas}.

Later, several enhancements to this scheme were considered. In particular, it was shown in Ref.\,\cite{Plick_NJP_12_083014_2010} that the performance of an SU(1,1) interferometer can be improved by seeding it with coherent light. In Ref.\,\cite{Anderson_Optica_4_752_2017}, various detection options for the SU(1,1) interferometer were considered and the new scheme of the truncated SU(1,1) interferometer, with the second OPA replaced by two homodyne detectors, was proposed. Review of various topologies and implementations of the SU(1,1) interferometers can be found in Refs.\,\cite{Chekhova_AOP_8_104_2016, Ou_APLPhot_5_080902_2020}.

\begin{figure}
  \includegraphics{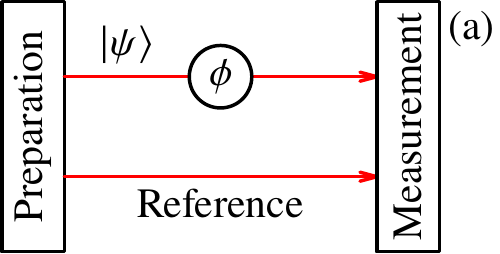}\quad
  \includegraphics{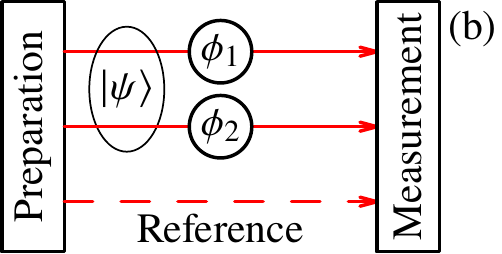}
  \caption{Conceptual schemes of the single-arm (a) and two-arm (b) interferometers. $\ket{\psi}$: single- or two-mode quantum state of the probing light.}\label{fig:3stage}
\end{figure}

The generalized conceptual schemes of the interferometric phase measurements, which encompasses both SU(2) and SU(1,1) options, are presented in Fig.\,\ref{fig:3stage}. Two configurations are shown: (a) with the unknown (signal) phase shift $\phi$ introduced into one arm only (the {\it single-arm} interferometer), and (b) with the two phase shifts $\phi_1$ and $\phi_2$ introduced respectively into each of the arms (the {\it two-arm} one). In both cases, the left block prepares the intitial single- or two-mode quantum state $\ket{\psi}$ of the probing light. Then the light experiences the phase shift(s) and after that is measured by the right block. In the single-arm case, evidently, a second beam providing the phase reference is necessary. It could correspond to the local oscillator beam of the homodyne detector in the SU(2) interferometer case or to the double-frequency pump for the second OPA in the SU(1,1) one. In the two-arm case, the reference beam is optional: it is required only if the sum phase shift $\phi_1+\phi_2$ has to be measured.

In literature, when analyzing the two-arms interferometers, the {\it asymmetric} case with the signal phase shift in one arm only:
\begin{equation}\label{phi_asymm}
  \phi_1 = \varphi \,,\quad \phi_2 = 0 \,.
\end{equation}
is often considered, implicitly assuming that it is equivalent to the {\it antisymmetric} one with the same value of the phase difference $\phi_1 - \phi_2$:
\begin{equation}\label{phi_antisymm}
  \phi_1 = \varphi/2 \,,\quad\varphi_2 = -\varphi/2 \,.
\end{equation}
Note also that the asymmetric layout looks identical to the single-arm interferometer, compare Fig\,\ref{fig:SU2}(b) for tha case of $\phi_2=0$ and Fig.\,\ref{fig:single}(a).

At the same time, it was shown in Refs.\,\cite{Jarzyna_PRA_85_011801_2012, Takeoka_PRA_96_052118_2017} that in general, the asymmetric and antisymmetrics layouts does not provide the same sensitivity, see details in subsection \ref{sec:phi_12}. Also,  it was shown in  Refs.\,\cite{Hofmann_PRA_79_033822_2009} that in the two-arms antisymmetric case the balanced configuration of the interferometer with the equal values of power in both arms is optimal. At the same time, in the single-arm case, the reference beam does not interact with the phase shifting object. Therefore, much more optical power can be sent into this beam, in comparison with the power in the probe beam, improving the sensitivity for a given optical power at the object (see discussion in the end of Sec.\,\ref{sec:QCRB_gen}). Due to these reasons, in this paper we consider the single-arm and two-arm cases separately.

Joint optimization of the preparation and measurement blocks could be a tedious task. In order to simplify it, an approach based on the {\it quantum Cram\`er-Rao bound} (QCRB) \cite{HelstromBook} is broadly used in the literature, see \eg reviews \cite{Demkowicz_PIO_60_345_2015, Anderson_ch35_2019, Ou_APLPhot_5_080902_2020} and the references therein. The QCRB defines the ultimate sensitivity, achievable for any given initial quantum state. Therefore, it allows to optimize the preparation block independently on the measurement block. Then, any measurement which saturate the QCRB  automatically will be an optimal one.

The QCRB could be understood in the following way. Start with the (hypothetical) Heisenberg uncertainty relation for the phase $\hat{\phi}$ and photons number $\hat{N}$ in the initial quantum state of the probing light \cite{Heitler1954}:
\begin{equation}\label{QCRB_1}
  \Delta N\Delta \phi \ge \frac{1}{2} \,.
\end{equation}
Evidently, the phase uncertainty $\Delta\phi$ defines the ultimate sensitivity limit achievable for a given quantum state of the probing light $\ket{\psi}$. In particular, in the coherent state case with $\Delta N = \sqrt{N}$ we obtain Eq.\,\eqref{SNL}. Squeezing of the phase quadrature  proportionally increases $\Delta N$, giving Eq.\,\eqref{SQZ}. The Heisenberg limit \eqref{HL} arises from the (strictly speaking, incorrect) argument that $\Delta N$ can not exceed $N$.

It is known however that a Hermitian phase operator $\hat{\phi}$, conjugate to the number of quanta operator $\hat{N}$, does not exist \cite{Carruthers_RMP_40_411_1968}. Therefore, the ``regular'' Heisenberg uncertainty relation of the form \eqref{QCRB_1} also does not exist. However, the QCRB for the particular case of the phase measurement can be expressed in the form identical to \eqref{QCRB_1},  see details in Sec.\,\ref{sec:QCRB}. In this case $\phi$ has the meaning of a $c$-number shift of the probing light phase, created by some external agent (for example, a displacement of the interferometer mirror(s)), and $\Delta N$ is still the photon number uncertainty of the probing light.

The goal of this review is to provide the {\it unified} view on the {\it practical} configurations of the quantum-enhanced interferometers, based on the QCRB. The ``unified'' here means that we analyze both the single- and two-arm interferometers and, for each case, both SU(2) and SU(1,1) preparation and measurement options. For all configurations, we optimize first the preparation block using the QCRB. Then we identify measurement procedures which saturate it.

The ``practical'' means the following limitations. (i) We focus on the high-precision measurements with $\Delta\phi\ll1$, because in the opposite case the required sensitivity can be easily obtained using the coherent state. (ii) We consider only the Gaussian quantum states, because they allow to saturate the fundamental limit \eqref{SQZ}, while technologies of their preparation (opposite to the exotic non-Gaussian states) are well developed and refined. (iii) We take into account that the useful amount of squeezing is limited by the optical losses in the interferometer by the condition \eqref{r_loss}. Assuming the realistic values of $1-\eta\gtrsim10^{-2}$, this means that the strong classical (coherent) pumping with
\begin{equation}\label{big_alpha}
  |\alpha|^2 \approx N \gg e^{2r} ,
\end{equation}
where $\alpha$ is the classical amplitude, is obligatory for the high-precision phase measurements. Due to this limitation, we do not consider the Heisenberg limit here. At the same time, in order to simplify the equations, we do not consider here the optical losses, assuming that
\begin{equation}
  e^r < e^{r_{\rm loss}} \,.
\end{equation}

This paper is organized as follows. In Sec.\,\ref{sec:QCRB} we reproduce derivation of the convenient for the phase sensitivity analysis form of the QCRB, which can be found in Ref.\,\cite{HelstromBook}, extending it to the multi-dimensional case.
Then in sections \ref{sec:single-arm} and \ref{sec:two-arm} we analyze the sensitivity of the single-arm and two-arm interferometers, respectively, following the formulated above  approach. In Sec.\,\ref{sec:conclusion} we summarize the main results of this paper.

Appendices contain all the calculations not so necessary for understanding of the main concepts discussed in this work.

\section{Quantum Cram\`er-Rao bound for phase measurements}\label{sec:QCRB}

\subsection{General case}\label{sec:QCRB_gen}

Let us consider some quantum system, \eg an optical field in the interferometer arms, in a state described by the density operator $\hat{\rho}$. Let $\hat{\rho}$ depend on $J\ge1$ parameters $\phi_1,\,\phi_2\,\dots,\phi_J$ which have to be estimated. Let
\begin{equation}
  \mathbb{B} = \|\mean{\delta\phi_j\,\delta\phi_k}\| \,,
\end{equation}
($j,k=1\dots J$) be the covariance matrix of the estimates $\tilde{\phi}_j$ of $\phi_j$, with
\begin{equation}
  \delta\phi_j = \tilde{\phi}_j - \phi_j \,.
\end{equation}

It was shown by C.\,Helstrom \cite{HelstromBook} that for any measurement procedure, $\mathbb{B}$ satisfies the following quantum Cram\`er-Rao inequality
\begin{equation}\label{QCRB_gen}
  \mathbb{B} \ge \mathbb{A}^{-1} \,,
\end{equation}
where
\begin{equation}
  \mathbb{A}
    = \biggl\|\frac{1}{2}\Tr[\hat{\rho}(\hat{L}_j\hat{L}_k + \hat{L}_k\hat{L}_j)]\biggr\|
    \label{bbA}
\end{equation}
is the {\it quantum Fisher information} matrix and $\hat{L}_j$ are the symmetrized logarithmic derivatives of $\hat{\rho}$, defined by
\begin{equation}\label{log_deriv0}
  \partd{\hat{\rho}}{\phi_j} = \frac{1}{2}(\hat{\rho}\hat{L}_j + \hat{L}_j\hat{\rho}) \,.
\end{equation}

Let now $\phi_1,\,\phi_2\,\dots$ be the phase shifts introduced in the respective arms of the interferometer. In this case,
\begin{equation}\label{QCRB_rho_a}
  \hat{\rho} = \hat{U}\hat{\rho}_0\hat{U}^\dag \,,
\end{equation}
where the density operator $\hat{\rho}_0$ corresponds to the optical field state before the phase shifts (\ie just after the first beamsplitter),
\begin{equation}\label{QCRB_rho_b}
  \hat{U} = \exp\biggl(-i\sum_{j=1}^J\hat{N}_j\phi_j\biggr)
\end{equation}
is the phase shifting evolution operator, and $\hat{N}_j$ are  the photon number operators of the respective arms of the interferometer. Note that they commute with each other:
\begin{equation}\label{N_commute}
  \forall j,k:\ [\hat{N}_j,\hat{N_k}] = 0 \,.
\end{equation}

In Ref.\,\cite{HelstromBook}, the particular case of this {\it displacement parameters estimation} problem for the single-dimensional case of $J=1$ was considered. It was shown that in this case, if the quantum state $\hat{\rho}$ is pure for all values of $\phi$,
\begin{equation}\label{QCRB_pure}
  \hat{\rho} = \ket{\psi}\bra{\psi} \,,
\end{equation}
then inequality \eqref{QCRB_gen} reduces to Eq.\,\eqref{QCRB_1}, with $\Delta N$ being the uncertainty of $\hat{N}$ in the quantum state $\ket{\psi}$.

In the commutative case of \eqref{N_commute}, this result can be trivially extended to the multi-parameter case. We show in App.\,\ref{app:Fisher} that in this case the Fisher information matrix is proportional to the variances matrix of the operators $\hat{N}_j$\footnote{Through the paper, we denote $\delta\hat{Q} = \hat{Q} -\mean{\hat{Q}}$ for any operator $\hat{Q}$.}:
\begin{equation}\label{Fisher_J}
  \mathbb{A} = 4\|\mean{\delta\hat{N}_j\delta\hat{N}_k}\| \,.
\end{equation}

Concluding this subsection, we would like to emphasize one actually trivial but important issue. The sensitivity limits \eqref{QCRB_1}, \eqref{QCRB_gen}, and therefore \eqref{SNL}\,-\,\eqref{HL} are defined in terms of the photon number statistics {\it at the phase shifting object(s)}, that is in the interferometer arm(s). At the same time, in literature they often are expressed in terms of the incident photon numbers. The latter  approach, while technically is correct, usually leads to more cumbersome final equations and sometimes misleading statements. For example, in Ref.\,\cite{Plick_NJP_12_083014_2010} the authors, considering the coherent light boosted SU(1,1) interferometer, claim that their scheme could provide sensitivity exceeding the HL, which is actually not the case in the strict sense of Eq.\,\eqref{HL}.

We would like to mention also again that in the most high-power contemporary interferometers, namely the laser interferometric gravitational-wave detectors, the optical power in the arms is limited not by the pump laser power, but by the nonlinear effects inside the interferometer. Therefore, in addition to the consistency with QCRB and to its mathematical cleanness, the approach based on the photon number statistics at the object(s) can be considered as more practical. Due to these reasons, we will follow it in this work.

\subsection{Differential and common modes}\label{sec:phi_pm}

Consider now the most relevant to interferometers particular case of $J=2$. In the vast majority of applications, the main goal of two-arm interferometers is not to measure two phase shifts $\phi_{1,2}$ in the arms independently, but to measure instead the differential phase shift
\begin{subequations}\label{phi_pm}
  \begin{equation}\label{phi_m}
    \phi_- = \frac{\phi_1 - \phi_2}{2} \,,
  \end{equation}
  In some cases, the common phase
  \begin{equation}\label{phi_p}
    \phi_+ = \frac{\phi_1 + \phi_2}{2} \,,
  \end{equation}
\end{subequations}
is also measured\footnote{Typically in literature another normalization $\phi_\pm = \phi_1\pm\phi_2$ is used, which enforces the factor $1/2$ in Eq.\,\eqref{N_pm} instead in order to preserve the condition \eqref{U_pm}, see \eg Refs.\cite{Jarzyna_PRA_85_011801_2012, Ataman_PRA_102_013704_2020}. We, following Ref.\,\cite{17a1MaKhCh}, prefer to use the normalization (\ref{phi_pm}, \ref{N_pm}), because it gives the final equations of the form more consistent between the single- and two-arm cases.}. In particular, in the laser gravitational-wave detectors this information is used for locking the common mode of the interferometer \cite{CQG_32_7_074001_2015}. Another interesting example is the experimental proposals aimed at the preparation of the entangled quantum state of two mirrors \cite{MuellerEbhardt2008, MuellerEbhardt2009, Schnabel_PRA_92_012126_2015}. However, due to vulnerability to various non-fundamental noise sources, in particular the laser technical noise, $\phi_+$ typically is measured with the precision significantly inferior to the one of $\phi_-$.

Therefore, it is convenient to rewrite the QCRB directly in terms of $\phi_\pm$. Introduce the common and differential values of the photon number in the interferometer:
\begin{equation}\label{N_pm}
  \hat{N}_\pm = \hat{N}_1 \pm \hat{N}_2 \,.
\end{equation}
Note that the evolution operator \eqref{QCRB_rho_b}, expressed in terms of $\phi_\pm$ and $\hat{N}_\pm$, retains its form:
\begin{equation}\label{U_pm}
  \hat{U} = e^{-i(\hat{N}_1\phi_1 + \hat{N}_2\phi_2)}
  = e^{-i(\hat{N}_+\phi_+ + \hat{N}_-\phi_-)} .
\end{equation}
Therefore, the QCRB for $\phi_\pm$ also has the form \eqref{QCRB_gen}, up to the evident modification of the notations (compare with Eqs.\,(4)\,-\,(6) of Ref.\,\cite{Lang_PRL_111_173601_2013}):
\begin{subequations}
  \begin{gather}
    \mathbb{B} = \smatrix{\mean{(\delta\phi_+)^2}}{\mean{\delta\phi_+\,\delta\phi_-}}
      {\mean{\delta\phi_+\,\delta\phi_-}}{\mean{(\delta\phi_-)^2}} , \\
    \mathbb{A}
      = 4\smatrix{\mean{(\delta\hat{N}_+)^2}}{\mean{\delta\hat{N}_+\,\delta\hat{N}_-}}
          {\mean{\delta\hat{N}_+\,\delta\hat{N}_-}}{\mean{(\delta\hat{N}_-)^2}} ,
  \end{gather}
\end{subequations}
where
\begin{equation}
  \delta\phi_\pm = \tilde{\phi}_\pm - \phi_\pm \,.
\end{equation}
The explicit form of the inverse Fisher information matrix can be readily found in this case:
\begin{equation}\label{inv_Fisher}
  \mathbb{A}^{-1} = \frac{1}{4\Det}
    \smatrix{\mean{(\delta\hat{N}_-)^2}}{-\mean{\delta\hat{N}_+\,\delta\hat{N}_-}}
      {-\mean{\delta\hat{N}_+\,\delta\hat{N}_-}}{\mean{(\delta\hat{N}_+)^2}} ,
\end{equation}
where
\begin{equation}
  \Det = \mean{(\delta\hat{N}_+)^2}\mean{(\delta\hat{N}_-)^2}
    - \mean{\delta\hat{N}_+\,\delta\hat{N}_-}^2 \,.
\end{equation}

\section{Single-arm interferometers}\label{sec:single-arm}

\begin{figure}
  \includegraphics{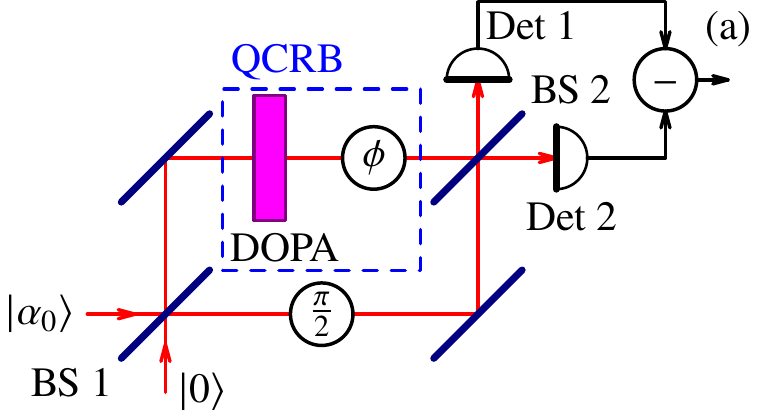}\quad
  \includegraphics{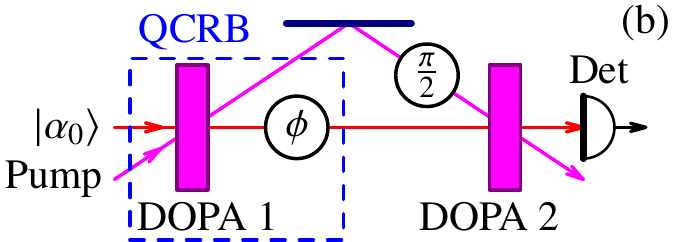}
  \caption{Single-arm SU(2) (a) and SU(1,1) (b) interfermeters. BS: beamsplitters; DOPA: degenerate optical parametric amplifiers; Det: detectors. The part, common for both schemes, is enclosed into the dashed rectangle and labeled as ``QCRB''.}\label{fig:single}
\end{figure}

\subsection{QCRB}\label{sec:QCRB_1}

Two configurations of the single-arm interferometers, the SU(2) and SU(1,1) ones, are shown in Figs.\,\ref{fig:single}(a) and (b), respectively. In both cases, we assume that the incident light is prepared in the coherent state. In the first case, the input beamsplitter BS~1 splits the incident beam into the probe and reference ones. The probe beam is squeezed by the degenerate optical parametric amplifier DOPA and then acquires the signal phase shift $\phi$. Finally, its phase quadrature is measured by the phase-sensitive balanced homodyne detector, consisting of the output beamsplitter BS~2 and two photodetectors Det~1 and Det~2. The second (reference) beam, with its phase shifted by $\pi/2$, serves as the local oscillator for the homodyne detector.

Note the unusual placement of the squeezer DOPA. In the case of the standard placement (in the ``south'' input port of BS~1), squeezing in the probe beam would be diluted by the vacuum fluctuations entering the interferometer through the classical pump (``west'') port, thus degrading the sensitivity. The placement of the squeezer directly in the signal arm solves this problem. However, as we show here, the optimal sensitivity of this scheme corresponds to the highly unbalanced regime with the BS~1 transmissivity $T\to1$. It is easy to see that in this case squeezer could be moved to its standard place without the sensitivity degradation.

In the SU(1,1) case, the incident beam goes directly to the squeezer DOPA~1, then acquires the signal phase shift $\phi$, and finally is anti-squeezed by the second parametric amplifier DOPA~2, excited by the same pump beam as the first one, but with the phase shifted by $\pi/2$ (which corresponds to inversion of the squeeze factor sign). In this case, the pump beam serves as the phase reference, allowing to use the simple direct detection instead of the homodyne one.

It is easy to see that these schemes differ only by the measurement methods, while the preparation and the phase shift parts which are relevant to the QCRB (in Fig.\,\ref{fig:single} they are enclosed into dashed rectangles and marked as "QCRB") are identical. Therefore, both schemes are characterized by the same Cram\`er-Rao limit \eqref{QCRB_1}, where $\Delta N$ is the uncertainty of the photon number $\hat{N}=\hat{a}^\dag\hat{a}$ in the squeezed coherent state generated by DOPA/DOPA~1 and $\hat{a}$, $\hat{a}^\dag$ are the corresponding annihilation and creation operators.

In the Heisenberg picture, the operator $\hat{a}$ can be expressed as follows:
\begin{equation}\label{b_1}
  \hat{a} = \alpha + \hat{z}\cosh r + \hat{z}^\dag e^{2i\theta}\sinh r \,,
\end{equation}
where  $\alpha$ is the classical amplitude of the probe beam at the phase shifting object, $\hat{z}$, $\hat{z}^\dag$ are the annihilation and creation operators of the input vacuum fluctuations of DOPA/DOPA~1,
$r>0$ is the squeeze factor and $\theta$ is the squeeze angle. We assume here without limiting the generality that $\arg\alpha=0$ (this assumption just defines the reference point for all phases).

It follows from Eq.\,\eqref{b_1} that
\begin{equation}
  \mean{(\delta\hat{N})^2}
  = \alpha^2(\cosh2r + \sinh2r\cos2\theta) + \frac{1}{2}\sinh^22r \,.
\end{equation}
The maximum of this expression is provided by
\begin{equation}\label{theta_0}
  \theta = 0 \,.
\end{equation}
This value of $\theta$ corresponds to anti-squeezing of the amplitude quadrature (the one which is in-phase with the classical carrier) and to the proportional squeezing of the phase quadrature. The resulting QCRB has the following form:
\begin{equation}\label{QCRB_single}
  \mean{(\delta\phi)^2} = \frac{1}{4\mean{(\delta\hat{N})^2}}
  = \frac{1}{4(\alpha^2e^{2r} + \frac{1}{2}\sinh^22r)} .
\end{equation}


\subsection{Homodyne detection}

Here we take into account explicitly the homodyne detector and the corresponding reference beam. It is well known that in the homodyne detection based schemes, the sensitivity {\it for a given optical power at the object} monotonously improves with the increase of the local oscillator power. Therefore, here we consider the general case of the unbalanced beamsplitter BS~1 with the power transmissivity $T\ne1/2$.

The annihilation operators of, respectively, the probe and the reference beams just before the output beamsplitter BS~2 (see Fig.\,\ref{fig:single}(a)) are the following:
\begin{subequations}\label{SU2_1_c}
  \begin{gather}
    \hat{c} = \hat{a}e^{-i\phi} \,, \label{c_1} \\
    \hat{c}_R = i\alpha_R + \hat{a}_R \,,
  \end{gather}
\end{subequations}
where $\alpha_R$, with $\arg\alpha_R=0$, is the classical amplitude of the reference beam, the factor $i$ takes into account the phase shift $\pi/2$ in the reference arm, and the annihilation operator $\hat{a}_R$ corresponds to a vacuum field.

The balanced output beamsplitter transforms them as follows:
\begin{equation}\label{SU2_1_d}
  \hat{d}_{1,2} = \frac{\hat{c} \pm \hat{c}_R}{\sqrt{2}} \,.
\end{equation}
Therefore, the differential number of quanta is equal to
\begin{equation}\label{SU2_1_n}
  \hat{n}_- = \hat{d}_1^\dag\hat{d}_1 - \hat{d}_2^\dag\hat{d}_2
  = \hat{c}^\dag\hat{c}_R + \hat{c}_R^\dag\hat{c} \,.
\end{equation}
Combining Eqs.\,\eqref{b_1}, \eqref{theta_0}, \eqref{SU2_1_c}\,-\,\eqref{SU2_1_n}, we obtain that the mean number and the variance of $\hat{n}_-$ are equal to, respectively,
\begin{subequations}\label{SU_2_1_stat}
  \begin{gather}
    \mean{\hat{n}_-} = -2\alpha\alpha_R\sin\phi \,, \\
    \mean{(\delta\hat{n}_-)^2} = \alpha^2 + \alpha_R^2(\cosh2r - \sinh2r\cos2\phi)
      + \sinh^2r \,.
  \end{gather}
\end{subequations}

Using the standard error propagation formula:
\begin{equation}
  \mean{(\delta\phi)^2} = \frac{\mean{(\delta\hat{n}_-)^2}}{G^2}\biggr|_{\phi\to0} \,,
\end{equation}
where
\begin{equation}
  G = \partd{\mean{\hat{n}_-}}{\phi} = 2\alpha\alpha_R\cos\phi
\end{equation}
is the gain factor, we obtain that
\begin{equation}
  \mean{(\delta\phi)^2} = \frac{1}{4}\biggl(
      \frac{e^{-2r}}{\alpha^2} + \frac{1}{\alpha_R^2} + \frac{\sinh^2r}{\alpha^2\alpha_R^2}
    \biggr) .
\end{equation}
The terms $\propto1/\alpha_R^2$ in this equation, which originate from the quantum noise of the reference beam, vanishe in the strongly asymmetric case of $\alpha_R^2 \to \infty$ (which corresponds to $\alpha_0\to\infty$, $T\to1$, and $\alpha_0\sqrt{1-T}=\alpha$), giving
\begin{equation}\label{QCRB_single1}
  \mean{(\delta\phi)^2} = \frac{e^{-2r}}{4\alpha^2} \,.
\end{equation}
This value differs from the fundamental limit \eqref{QCRB_single} only by absence of the small term $\frac{1}{2}\sinh^22r$ in the denominator.

It was shown in Ref.\,\cite{Pezze_PRL_100_073601_2008}, that this additional term arises because the mean value of $\hat{n}_-$ does not contain full information about $\phi$. Some of this information is contained in the higher momenta of $\hat{n}_-$ and can be recovered using more sophisticated multi-shot Bayesian measurements, as it was proposed in \cite{Pezze_PRL_100_073601_2008}. However, in the case of \eqref{big_alpha}, the ordinary homodyne measurement is sufficient.

\subsection{SU(1,1) measurement}\label{sec:SU11_1}

Consider now the single-arm SU(1,1) interferometer, shown in Fig.\,\ref{fig:single}(b). In this case, the phase dependence of the output photon number $\hat{n}=\hat{d}^\dag\hat{d}$, where $\hat{d}$, $d^\dag$ are the corresponding annihilation and creation operators, is created by the DOPA~2, which preforms the anti-squeezing operation
\begin{equation}\label{d_SU11}
  \hat{d} = \hat{c}\cosh R - \hat{c}^\dag e^{2i\theta}\sinh R \,.
\end{equation}
Here $\hat{c}$ is given by Eq.\,\eqref{c_1}, $R>0$ is the anti-squeeze factor, and $\theta$ is th same squeeze angle as in Eq.\,\eqref{b_1}.

The mean value and variance of $\hat{n}$ are calculated in App.\ref{app:SU_1_1}, see Eqs.\,\eqref{n_SU_1_1}. Here we consider only the particular case of
\begin{equation}\label{big_R}
  e^{2R} \gg 1 \,,
\end{equation}
because (i) it significantly simplify the equaitons, while providing the sensitivity approaching the QCRB \eqref{QCRB_single}, and (ii) allows to suppress influence of the photodetector inefficiency \cite{17a1MaKhCh} (note that the perfectrly realistic case of the 10\,dB squeezing, $e^{2R}=10\gg1$). In this case,
\begin{subequations}\label{SU_1_1_stat}
  \begin{gather}
    \mean{\hat{n}}
      = e^{2R}\biggl[\alpha^2\sin^2(\theta+\phi) + \frac{|\sigma|^2}{4}\biggr] , \\
    \mean{(\delta\hat{n})^2}
      = e^{4R}|\sigma|^2\biggl[\alpha^2\sin^2(\theta+\phi) + \frac{|\sigma|^2}{8}\biggr] \,,
  \end{gather}
\end{subequations}
where
\begin{equation}
  |\sigma|^2 = \cosh2r - \sinh2r\cos2\phi \,.
\end{equation}

Using again the error propagation formula:
\begin{equation}
  \mean{(\delta\phi)^2} = \frac{\mean{(\delta\hat{n})^2}}{G^2}\biggr|_{\phi\to0} \,,
\end{equation}
with
\begin{equation}
  G = \partd{\mean{\hat{n}}}{\phi}
  = e^{2R}\biggl[\alpha^2\sin2(\theta+\phi) + \frac{1}{2}\sinh2r\sin2\phi\biggr],
\end{equation}
we obtain that
\begin{equation}
  \mean{(\delta\phi)^2} = \frac{\mean{(\delta\hat{n})^2}}{G^2}\biggr|_{\phi\to0}
  = \frac{e^{-2r}}{4\alpha^2\cos^2\theta}
      \biggl(1 + \frac{e^{-2r}}{8\alpha^2\sin^2\theta}\biggr) .
\end{equation}
In the case of a small, but not non-zero squeeze angle:
\begin{equation}
  \frac{e^{-2r}}{N} \ll \theta^2 \ll 1 \,,
\end{equation}
we again come to Eq.\,\eqref{QCRB_single1}.

\section{Two-arm interferometers}\label{sec:two-arm}

\subsection{QCRB}\label{sec:QCRB_2}

\begin{figure}
  \includegraphics{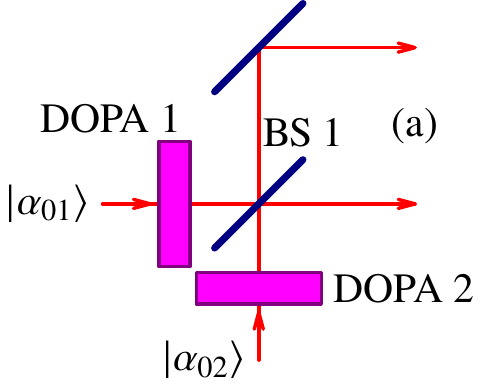}\quad
  \includegraphics{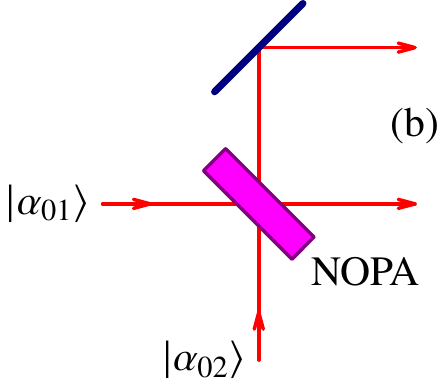}
  \caption{Preparation options for the two-arm interferometer shown in Fig.\,\ref{fig:3stage}(b). (a): SU(2) interferometer; (b): SU(1,1) interferometer. BS~1: beamsplitters; DOPA~1,2: degenerate optical parametric amplifiers; NOPA: non-degenerate optical parametric amplifier.}\label{fig:prep}
\end{figure}


Two variants of the preparation block of the two-arm interferometer (see Fig.\,\ref{fig:3stage}(b)) are shown in Fig.\,\ref{fig:prep}: the ``SU(2)'' (a) and the ``SU(1,1)'' (b) ones. Let us start with the first one.

We consider the most general case, assuming that the both input ports of the beamsplitter BS~1 are excited by two squeezed coherent states generated by two degenerate parametric amplifiers DOPA~1,2. The annihilation operators describing these quantum states in the Heisenberg picture can be presented as follows ($j=1,2$):
\begin{equation}\label{a12_SU2}
  \hat{a}_j = \alpha_j + \hat{z}_j\cosh r_j
    + \hat{z}_j^\dag e^{2i\theta_j}\sinh r_j \,,
\end{equation}
where $\alpha_j$ are the classical amplitudes, the operators $\hat{z}_j$ correspond to vacuum fields, $r_j$ are the squeeze factors and $\theta_j$ are the squeeze angles. After the input beamsplitter BS~1, we obtain correspondingly:
\begin{equation}\label{b12_SU2}
  \hat{b}_1 = \frac{\hat{a}_1 + \hat{a}_2}{\sqrt{2}} \,, \quad
  \hat{b}_2 = \frac{\hat{a}_1 - \hat{a}_2}{\sqrt{2}} \,,
\end{equation}

In particular, the classical amplitudes of the beamsplitter output beams are equal to
\begin{equation}\label{beta12_SU2}
  \beta_1 = \frac{\alpha_1 + \alpha_2}{\sqrt{2}} \,, \quad
  \beta_2 = \frac{\alpha_1 - \alpha_2}{\sqrt{2}} \,.
\end{equation}
Taking into account that the best sensitivity for the differential phase $\phi_-$ is provided by the balanced interferometer \cite{Hofmann_PRA_79_033822_2009}, we suppose that $|\beta_1| = |\beta_2|$. Then, without further limiting the generality (and setting thus the reference point for all phases), we assume that
\begin{equation}\label{beta_12}
  \beta_{1,2} = \frac{\alpha e^{\pm i\zeta}}{\sqrt{2}} \,,
\end{equation}
where $\arg\alpha=0$ and $2\zeta$ is the relative phase shift between $\beta_1$ and $\beta_2$. It follows from Eqs.\,\eqref{beta12_SU2}, that in this case,
\begin{equation}\label{alpha12}
  \alpha_1 = \alpha\cos\zeta \,, \quad \alpha_2 = i\alpha\sin\zeta
\end{equation}
and
\begin{equation}
  |\beta_1|^2 + |\beta_2|^2 = |\alpha_1|^2 + |\alpha_2|^2 = |\alpha|^2 \,.
\end{equation}

It follows from Eqs.\,\eqref{b12_SU2}, that the common and differential values of the photon numbers at the phase shifting objects are equal to
\begin{subequations}
  \begin{gather}
    \hat{N}_+ = \hat{b}_1^\dag\hat{b}_1 + \hat{b}_2^\dag\hat{b}_2
      = \hat{a}_1^\dag\hat{a}_1 + \hat{a}_2^\dag\hat{a}_2 \,, \\
    \hat{N}_- = \hat{b}_1^\dag\hat{b}_1 - \hat{b}_2^\dag\hat{b}_2
      = \hat{a}_1^\dag\hat{a}_2 + \hat{a}_2^\dag\hat{a}_1 \,.
  \end{gather}
\end{subequations}
The second momenta of these operators are calculated in App.\,\ref{app:dN_pm_SU2}. In the particular case of \eqref{alpha12}, they can be presented as follows:
\begin{subequations}\label{DN_pm_raw}
  \begin{multline}
    \mean{(\delta\hat{N}_+)^2} = \alpha^2[
        (\cosh2r_1 + \sinh2r_1\cos2\theta_1)\cos^2\zeta
        + (\cosh2r_2 - \sinh2r_2\cos2\theta_2)\sin^2\zeta
      ] \\
      + \frac{1}{2}(\sinh^22r_1 + \sinh^22r_2) \,, \label{DN_pm(a)}
  \end{multline}
  \begin{multline}
    \mean{(\delta\hat{N}_-)^2} = \alpha^2[
        (\cosh2r_1 - \sinh2r_1\cos2\theta_1)\sin^2\zeta
        + (\cosh2r_2 + \sinh2r_2\cos2\theta_2)\cos^2\zeta
      ] \\
      + \sinh^2(r_1+r_2)\cos^2(\theta_1-\theta_2) + \sinh^2(r_1-r_2)\sin^2(\theta_1-\theta_2) \,, \label{DN_pm(b)}
  \end{multline}
  \begin{equation}
    \mean{\delta\hat{N}_+\delta\hat{N}_-}
      = \frac{\alpha^2}{2}(\sinh2r_1\sin2\theta_1 + \sinh2r_2\sin2\theta_2)\sin2\zeta \,.
  \end{equation}
\end{subequations}
Consider now three characteristic scenarios.

\paragraph{Single squeezer.}

Start with the case of a single squeezer. Just to be specific, we suppose that
\begin{equation}\label{single_squeezer_a}
  r_1 = 0 \,, \quad r_2 > 0\,.
\end{equation}
It follows from Eqs.\,\eqref{DN_pm_raw}, that in this case,
\begin{subequations}\label{DN_pm_single}
  \begin{gather}
    \mean{(\delta\hat{N}_+)^2}
      = \alpha^2[\cos^2\zeta + (\cosh2r_2 - \sinh2r_2\cos2\theta_2)\sin^2\zeta]
        + \frac{1}{2}\sinh^22r_2   \,, \\
    \mean{(\delta\hat{N}_-)^2}
      = \alpha^2[\sin^2\zeta + (\cosh2r_2 + \sinh2r_2\cos2\theta_2)\cos^2\zeta] + \sinh^2r_2 \,,
      \label{DN_pm_single(b)}\\
    \mean{\delta\hat{N}_+\delta\hat{N}_-}
      = \frac{\alpha^2}{2}\sinh2r_2\sin2\theta_2\sin2\zeta \,.  \label{DN_pm_single(c)}
  \end{gather}
\end{subequations}
Then recall that the measurement of $\phi_-$ has higher priority than the measurement of $\phi_+$. Therefore, evidently, the following parameters have to be used, which maximize the leading (proportional to $\alpha^2$) term in Eq.\,\eqref{DN_pm_single(b)}:
\begin{equation}\label{single_squeezer_b}
  \theta_2 = 0 \,,\quad \zeta = 0 \,.
\end{equation}
In this case it follows from Eq.\,\eqref{inv_Fisher} that
\begin{subequations}\label{DN_pm_Caves}
  \begin{gather}
    \mean{(\delta\phi_+)^2} = \frac{1}{4\bigl(\alpha^2 + \frac{1}{2}\sinh^22r_2\bigr)} \,, \\
    \mean{(\delta\phi_-)^2} = \frac{1}{4(\alpha^2e^{2r_2} + \sinh^2r_2)} \,,
  \end{gather}
\end{subequations}
\begin{equation}
  \mean{\delta\phi_+\delta\phi_-} = 0 \label{DN_pm_Caves(c)} \,,
\end{equation}
Note that conditions \eqref{single_squeezer_a}, \eqref{single_squeezer_b} describe injection of the classical carrier into only one (bright) port ($\alpha_2=0$) and injection of the squeezed vacuum into the second (dark) one. This regime exactly corresponds the canonical case of \cite{Caves1981}.

Note also that due to the condition \eqref{DN_pm_Caves(c)}, the measurement imprecision of $\phi_-$ is disentangled from the one of $\phi_+$, which is typically contaminated by technical noise. This useful feature is true also for the "two squeezers" and "SU(1,1) type preparation" cases considered below. For brevity, we will not mention it again and will omit the corresponding equations for $\mean{\delta\phi_+\delta\phi_-}$.

\paragraph{Two squeezers.}

Suppose now that the both input ports are squeezed. It is natural to assume that both squeeze factors are limited by the same technological constraints and therefore $|r_1|\sim|r_2|$. In this case, the leading (proportional to $\alpha^2$) terms in  Eqs.\,\eqref{DN_pm(a)}, \eqref{DN_pm(b)} can be maximized simultaneously by either
\begin{subequations}\label{two_arm_opt}
  \begin{equation}\label{two_arm_opt_1}
    r_1 \sim r_2 > 0 \,, \quad \theta_1 = \theta_2 = 0 \,, \quad \zeta = 0 \,,
  \end{equation}
  or
  \begin{equation}\label{two_arm_opt_2}
    r_1 \sim r_2 > 0 \,, \quad \theta_1 = \theta_2 = \frac{\pi}{2} \,, \quad
      \zeta = \frac{\pi}{2} \,.
  \end{equation}
\end{subequations}
Both these options actually are equivalent, up to renumbering of the ports. Similar to the previous (single squeezer) case, they also correspond to injection bright carrier into one of the ports and injection of the squeezed vacuum into the second one. But this time, the light at the bright port is also squeezed.

To be specific, we consider the first option, which gives:
\begin{subequations}\label{DN_pm_double}
  \begin{gather}
    \mean{(\delta\phi_+)^2}
      = \frac{1}{4[\alpha^2e^{2r_1} + \frac{1}{2}(\sinh^22r_1 + \sinh^22r_2)]} , \\
    \mean{(\delta\phi_-)^2} = \frac{1}{4[\alpha^2e^{2r_2} + \sinh^2(r_1+r_2)]}
      \label{DN_pm_double(b)}.
  \end{gather}
\end{subequations}
Therefore, the bright port squeezer allows to obtain the sub-SNL sensitivity also for the common phase.

\paragraph{SU(1,1)-type preparation.}

Finally, consider the input part of the SU(1,1) interferometer, see Fig.\,\ref{fig:prep}(b). In this case, the incident fields at the phase shifting objects can be presented as follows:
\begin{subequations}\label{b12_SU11}
	\begin{gather}
  	\hat{b}_1 = \beta_1 + \hat{z}_+\cosh r + \hat{z}_-^\dag e^{2i\theta}\sinh r \,, \\
  	\hat{b}_2 = \beta_2 + \hat{z}_-\cosh r + \hat{z}_+^\dag e^{2i\theta}\sinh r \,,
	\end{gather}
\end{subequations}
where $\beta_j$ are still given by Eqs.\,\eqref{beta_12} and $\hat{z}_\pm$ correspond to the vacuum fields. It is easy to see, that up the trivial substitution
\begin{equation}
  \hat{z}_\pm \to \frac{\hat{z}_1 \pm \hat{z}_2}{\sqrt{2}} \,,
\end{equation}
Eqs.\,\eqref{b12_SU11} coincide with Eqs.\,\eqref{a12_SU2}, \eqref{b12_SU2} for the particular case of
\begin{subequations}\label{SU11-SU2}
  \begin{gather}
  -r_1 = r_2 = r \,, \label{SU11-SU2(a)} \\
    \theta_1 = \theta_2 = \theta \,.
  \end{gather}
\end{subequations}
This means that the SU(1,1) type scheme of Fig.\,\ref{fig:prep}(b) can be treated as a special case of the SU(2) scheme of Fig.\,\ref{fig:prep}(a) with the particular choice of the squeeze factors given by \eqref{SU11-SU2}. Therefore, we can reuse Eqs.\,\eqref{DN_pm_raw}. Combining them with Eqs.\,\eqref{SU11-SU2}, we obtain that
\begin{subequations}\label{DN_pm_SU11_raw}
  \begin{gather}
    \mean{(\delta\hat{N}_+)^2} = \alpha^2(\cosh2r - \sinh2r\cos2\theta) + \sinh^22r \,, \\
    \mean{(\delta\hat{N}_-)^2} = \alpha^2(\cosh2r + \sinh2r\cos2\theta) \,.
  \end{gather}
\end{subequations}
independently  on $\zeta$.

In this case, trade-off is possible between the $\phi_+$ and $\phi_-$. In particular, if
$r>0$ and $\theta=0$, then we have the sensitivity overcoming the SNL for $\phi_-$, but proportionally degraded for $\phi_+$:
\begin{subequations}\label{DN_pm_SU11(1)}
  \begin{gather}
    \mean{(\delta\phi_+)^2} = \frac{1}{4(\alpha^2e^{-2r} + \sinh^22r)} , \\
    \mean{(\delta\phi_-)^2} = \frac{1}{4\alpha^2e^{2r}} .
  \end{gather}
\end{subequations}
If $r>0$ and $\theta=\pi/2$, then the situation is opposite:
\begin{subequations}\label{DN_pm_SU11(2)}
  \begin{gather}
    \mean{(\delta\phi_+)^2} = \frac{1}{4(\alpha^2e^{2r} + \sinh^22r)} , \\
    \mean{(\delta\phi_-)^2} = \frac{1}{4\alpha^2e^{-2r}} \,.
  \end{gather}
\end{subequations}
However, the latter case hardly have any practical sense.

The following interesting case is also worth mentioning:
\begin{equation}
  \theta = \frac{\pi}{4} \,,
\end{equation}
which, similar to the two squeezers case, see Eqs.\,\eqref{DN_pm_double}, also allows to overcome the SNL for both $\phi_+$ and $\phi_-$ simultaneously, but to a lesser degree:
\begin{subequations}\label{DN_pm_SU11(3)}
  \begin{gather}
    \mean{(\delta\phi_+)^2} = \frac{1}{4(\alpha^2\cosh2r + \sinh^22r)} , \\
    \mean{(\delta\phi_-)^2} = \frac{1}{4\alpha^2\cosh2r} .
  \end{gather}
\end{subequations}


\subsection{Measurement of individual phase shifts in the arms}\label{sec:phi_12}

Here we analyze the measurement precision $\mean{(\delta\phi_{1,2})^2}$ for the individual phase shifts in the arms
\begin{equation}\label{phi_pm12}
  \phi_{1,2} = \phi_+ \pm \phi_- \,.
\end{equation}
In general, in order to recover them, both $\phi_+$ and $\phi_-$ have to be measured. The measurement of $\phi_+$, in turn, requires a detection scheme which uses an additional (third) phase reference beam, \eg the homodyne detector.

The values of measurement errors $\mean{(\delta\phi_{1,2})^2}$ crucially depend on a priori information on $\phi_{1,2}$. It follows form Eqs.\,(\ref{DN_pm_Caves(c)}, \ref{phi_pm12}) that in the absence of this information, the mean square measurement errors of $\phi_+$ and $\phi_-$ add up:
\begin{equation}
  \mean{(\delta\phi_{1,2})^2} = \mean{(\delta\phi_+)^2} + \mean{(\delta\phi_-)^2} \,,
\end{equation}
where $\mean{(\delta\phi_\pm)^2}$ are given by Eqs.\,\eqref{DN_pm_Caves}, \eqref{DN_pm_double}, or \eqref{DN_pm_SU11(1)}, depending on the input state prepration procedure.

At the same time, suppose that an unknown phase shift $\varphi$ is introduced into one, say the first, arm only and the phase shift in another arm is known --- say, equal to zero, see Eqs.\,\eqref{phi_asymm}. It follows from Eqs.\,\eqref{phi_pm}, that in this asymmetric case,
\begin{equation}
  \phi_+ = \phi_- = \frac{\varphi}{2} \,.
\end{equation}
This means that both the ``$+$'' and the ``$-$'' channels carry the same information on $\varphi$. It is easy to show that in this case, with account for the condition \eqref{DN_pm_Caves(c)}, the Fisher information values for both channel are summed:
\begin{equation}\label{dvarphi2}
  \mean{(\delta\varphi)^2}
  = \frac{4}{1/\mean{(\delta\phi_+)^2} + 1/\mean{(\delta\phi_-)^2}} \,.
\end{equation}

Another important particular case is the antisymmetric distribution of the phase shifts, see Eqs.\,\eqref{phi_antisymm}. It corresponds to
\begin{equation}
  \phi_+ = 0 \,, \quad \phi_- = \frac{\varphi}{2} \,.
\end{equation}
In this case, the ``+'' channel does not provide any information on $\varphi$; correspondingly,
\begin{equation}\label{dvarphi1}
  \mean{(\delta\varphi)^2} = 4\mean{(\delta\phi_-)^2} \,.
\end{equation}
Evidently, the additional phase reference is useless in this case.

It was shown in Ref.\,\cite{Jarzyna_PRA_85_011801_2012}, that the asymmetric and antisymmetric cases are equivalent only within the class of detection schemes insensitive to the common phase. Eqs.\,\eqref{dvarphi2} and \eqref{dvarphi1} clearly show the reason for this. Really, the first one contains the Fisher information from the ``+'' channel, in addition to the ``--'' one. Taking into account, that in the double squeezed input case there could be $\mean{(\delta\phi_+)^2} \approx \mean{(\delta\phi_-)^2}$, the value of \eqref{dvarphi2} could be twice as small as of \eqref{dvarphi1}. At the same time, consider the asymmetric case with the neglected common phase information (assuming for example, that a detection scheme which does not employ the reference beam is used). In this case, putting in Eq.\,\eqref{dvarphi2} $\mean{(\delta\phi_+)^2}\to\infty$, we obtain Eq.\,\eqref{dvarphi1}.

It was shown also in Ref.\,\cite{Takeoka_PRA_96_052118_2017} that in the antisymmetric case, if the vacuum state is sent into one of the input ports of the two-arm interferometer, then, independently on the quantum state at the other port, the sensitivity can not exceed the SNL; however, in the asymmetric case this no-go theorem does not hold. The origin of this peculiar limitation also can be seen from Eqs.\,\eqref{dvarphi2} and \eqref{dvarphi1}.

Really, consider Eqs.\,\eqref{DN_pm_double}, assuming that $r_2=0$, which corresponds to the vacuum state at the second input port. In this case, $\phi_-$ can not be measured with the precision better than the SNL, but $\phi_+$ still can. Due to this reason, in the antisymmetric case the precision is limited by the SNL, see Eq.\,\eqref{dvarphi1}. But the asymmetric case, the SNL can be overcome due to the Fisher information, provided by the ``+'' channel, see  Eq.\,\eqref{dvarphi2}.

\begin{figure*}
  \includegraphics{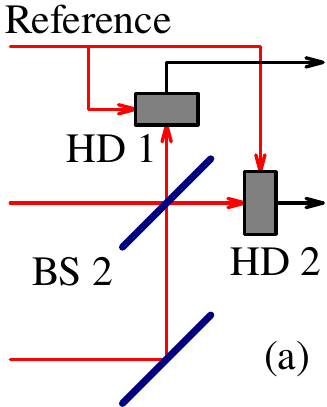}\quad
  \includegraphics{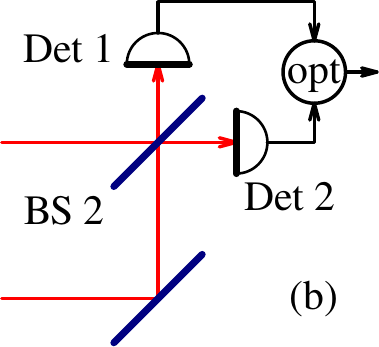}\quad
  \includegraphics{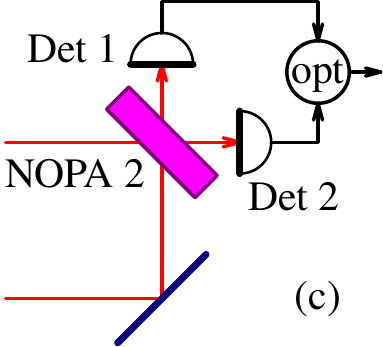}
  \caption{Measurement options for the two-arm interferometer shown in Fig.\,\ref{fig:3stage}(b). (a): Double homodyne detection; (b): double direct detection;
  (c): SU(1,1) type measurement. BS~2: beamsplitter; NOPA~2: non-degenerate optical parametric amplifier; HD~1,2: homodyne detectors; Det~1,2: detectors; ``opt'' means the optimally weighted sum of the two outputs.}\label{fig:meas}
\end{figure*}

\subsection{Double homodyne detection}\label{sec:homodyne2}

In this and the next two subsections, we consider three variants of the measurement block of the two-arm interferometer (see Fig.\,\ref{fig:3stage}(b)), shown in Fig.\,\ref{fig:meas}: the double homodyne detection (a),  the double direct detection (b), and the SU(1,1) type measurement (c). Here we start with the first one.

The annihilation operators of the optical fields after introducing the signal phase shifts are the following ($j=1,2$):
\begin{equation}\label{c12_SU2}
  \hat{c}_j = \hat{b}_je^{-i\phi_j} \,.
\end{equation}
The output beamsplitter BS~2 modifies them as follows:
\begin{equation}\label{d12_SU2}
  \hat{d}_1 = \frac{\hat{c}_1 + \hat{c}_2}{\sqrt{2}} \,, \quad
  \hat{d}_2 = \frac{\hat{c}_1 - \hat{c}_2}{\sqrt{2}} \,.
\end{equation}
Combining Eqs.\,\eqref{b12_SU2}, \eqref{c12_SU2}, \eqref{d12_SU2}, we obtain that
\begin{subequations}\label{d12_SU2_1}
  \begin{gather}
    \hat{d}_1 = (\hat{a}_1\cos\phi_- - i\hat{a}_2\sin\phi_-)e^{-i\phi_+} \,, \\
    \hat{d}_2 = (-i\hat{a}_1\sin\phi_- + \hat{a}_2\cos\phi_-)e^{-i\phi_+} \,.
  \end{gather}
\end{subequations}

Then, following the results of Sec.\,\ref{sec:QCRB_2}, we assume that
\begin{subequations}\label{a12_SU2meas}
  \begin{gather}
    \hat{a}_1 = \alpha + \hat{z}_1\cosh r_1 + \hat{z}_1^\dag\sinh r_1 \,, \\
    \hat{a}_2 = \hat{z}_2\cosh r_2 + \hat{z}_2^\dag\sinh r_2 \,,
  \end{gather}
\end{subequations}
where $\arg\alpha = 0$. Combining Eqs.\,(\ref{d12_SU2_1}, \ref{a12_SU2meas}) and taking into account that $|\alpha|\gg1$ and $|\phi_\pm|\ll1$, we obtain finally:
\begin{subequations}\label{d12_SU2meas}
  \begin{gather}
    \hat{d}_1 = \alpha(1-i\phi_+) + \hat{z}_1\cosh r_1 + \hat{z}_1^\dag\sinh r_1 \,, \\
    \hat{d}_2 = -i\alpha\phi_- + \hat{z}_2\cosh r_2 + \hat{z}_2^\dag\sinh r_2 \,.
  \end{gather}
\end{subequations}
It follows from these equations, that the first (bright) output contains information about the common phase $\phi_+$, and the second (dark) one --- about the differential one $\phi_-$. Both could be measured independently by homodyne detectors.

In order to calculate the measurement errors, it is convenient to introduce the sine quadratures of the input and output fields ($j=1,2$):
\begin{equation}
  \hat{z}_j^s = \frac{\hat{z}_j - \hat{z}_j^\dag}{i\sqrt{2}} \,, \quad
  \hat{d}_j^s = \frac{\hat{d}_j - \hat{d}_j^\dag}{i\sqrt{2}} \,.
\end{equation}
In these notations, Eqs.\,\eqref{d12_SU2meas} give:
\begin{subequations}\label{d12s_SU2meas}
  \begin{gather}
    \hat{d}_1^s = -\sqrt{2}\alpha\phi_+ + \hat{z}_1^se^{-r_1} \,, \\
    \hat{d}_2^s = -\sqrt{2}\alpha\phi_- + \hat{z}_2^se^{-r_2} \,.
  \end{gather}
\end{subequations}
Taking into account that the variances of $\hat{z}_{1,2}^s$ are equal to $1/2$, we obtain:
\begin{subequations}\label{dphi_pm_HD}
  \begin{gather}
    \mean{(\delta \phi_+)^2} = \frac{e^{-2r_1}}{4\alpha^2} \,, \\
    \mean{(\delta \phi_-)^2} = \frac{e^{-2r_2}}{4\alpha^2} \,.
  \end{gather}
\end{subequations}
In the asymptotic case of \eqref{big_alpha}, these results are in the full agreement with the ones predicted by the QCRB, see Eqs.\,\eqref{DN_pm_Caves}, \eqref{DN_pm_double}, \eqref{DN_pm_SU11(1)}.

\subsection{Double direct detection}\label{sec:direct2}

Consider now the double direct detection case, shown in Fig.\,\ref{fig:meas}(b).
The photon numbers at the outputs of the two photodetectors can be presented as follows:
\begin{subequations}
  \begin{gather}
    \hat{n}_1^{\rm out} = \hat{d}_1^\dag\hat{d}_1
      = \hat{n}_1\cos^2\phi_- + \hat{n}_2\sin^2\phi_- + \hat{Y}\cos\phi_-\sin\phi_- \,, \\
    \hat{n}_2^{\rm out} = \hat{d}_2^\dag\hat{d}_2
      = \hat{n}_1\sin^2\phi_- + \hat{n}_2\cos^2\phi_- - \hat{Y}\cos\phi_-\sin\phi_- \,,
  \end{gather}
\end{subequations}
where ($j=1,2$)
\begin{equation}
  \hat{n}_j = \hat{a}_j^\dag\hat{a}_j
\end{equation}
are the incident photon numbers and
\begin{equation}
  \hat{Y} = i(\hat{a}_2^\dag\hat{a}_1 - \hat{a}_1^\dag\hat{a}_2) \,.
\end{equation}
Correspondingly, the common and differential output photon numbers are equal to:
\begin{subequations}\label{n_pm_out}
  \begin{gather}
    \hat{n}_+^{\rm out} = \hat{n}_1^{\rm out} + \hat{n}_2^{\rm out} = \hat{n}_+  \,, \\
    \hat{n}_-^{\rm out} = \hat{n}_1^{\rm out} - \hat{n}_2^{\rm out}
      = \hat{n}_-\cos2\phi_- + \hat{Y}\sin2\phi_- \,.
  \end{gather}
\end{subequations}
where
\begin{equation}\label{n_pm}
  \hat{n}_\pm = \hat{n}_1 \pm \hat{n}_2 \,.
\end{equation}

It is evident that in this case, due to the absence of the common phase reference, only the differential phase shift $\phi_-$ can be measured. Note also that $\hat{n}_+^{\rm out}$ does not depend on $\phi_-$. Due to this reason, in literature typically the measurement of the differential photons numbers $\hat{n}_-^{\rm out}$ only is considered, see \eg Ref.\,\cite{Anderson_ch35_2019}. However, fluctuations of $\hat{n}_+^{\rm out}$ and $\hat{n}_-^{\rm out}$ correlate. Therefore, the measurement  of $\hat{n}_+^{\rm out}$ provides some information about the noise of $\hat{n}_-^{\rm out}$, and taking this information into account by means of the optimal combination of both outputs, it is possible to reduce the phase measurement error \cite{20a1ShSaFrMiChKh}. The resulting measurement error can be presented as follows:
\begin{equation}\label{dphi_direct_raw}
  \mean{(\delta\phi_-)^2} = \frac{1}{G^2}\biggl[
      \mean{(\delta\hat{n}_-^{\rm out})^2}
      - \frac{\mean{\delta\hat{n}_-^{\rm out}\delta\hat{n}_+^{\rm out}}^2}
          {\mean{(\delta\hat{n}_+^{\rm out})^2}}
    \biggr] ,
\end{equation}
where
\begin{equation}\label{G_direct_1}
  G = \partd{\mean{\hat{n}_-^{\rm out}}}{\phi_-}
\end{equation}
is the gain factor.

Here we, similarly to the double homodyne detection case (see Sec.\,\ref{sec:homodyne2}), follow the results of Sec.\,\ref{sec:QCRB_2}, assuming that the incident fields are described by Eqs.\,\eqref{a12_SU2meas}. The explicit form of Eq.\,\eqref{dphi_direct_raw} for this case is calculated in App.\,\ref{app:direct2}, see Eq.\,\eqref{dphi_direct_gen}.

Note that the squeeze factor $r_1$ (which appears implicitly in these equations) is responsible for the common phase $\phi_+$ sensitivity, see Sec.\,\ref{sec:QCRB_2}. Therefore, the ``two squeezers'' option has no sense in the direct detection case, which is insensitive to $\phi_+$. Consider the two remaining ones. In the single squeezer case of Eq.\,\eqref{single_squeezer_a},  Eq.\,\eqref{dphi_direct_gen} simplifies to
\begin{equation}\label{dphi_direct_single}
  \mean{(\delta\phi_-)^2} = \frac{1}{4(\alpha^2 - \sinh^2r)^2}\biggl(
      \alpha^2e^{-2r} + \sinh^2r
      + \frac{2\alpha^2\sinh^22r}{\alpha^2 + \sinh^22r}\cot^22\phi_-
    \biggr) ,
\end{equation}
In the SU(1,1)-type preparation case, see Eq.\,\eqref{SU11-SU2(a)}, we obtain:
\begin{equation}\label{dphi_direct_SU11}
  \mean{(\delta\phi_-)^2} = \frac{1}{4(\alpha^2e^{-2r} + \sinh^2r)}\biggl(
      e^{-4r} + \frac{2\alpha^2\sinh^22r + \sinh^42r}{\alpha^4\sin^22\phi_-}
    \biggr) .
\end{equation}

In both cases, the last terms in the parentheses can be canceled by introducing a sufficiently big given constant displacement into $\phi_-$. The remaining parts, in the asymptotical case of \eqref{big_alpha}, coincide with QCRB.

\subsection{SU(1,1) measurement}\label{sec:SU11_meas}

Finally, consider the SU(1,1) measurement case, shown in Fig.\,\ref{fig:meas}(b). In this case, with account for Eqs.\,\eqref{c12_SU2}, the optical fields after the NOPA~2 can be presented as follows:
\begin{subequations}
  \begin{gather}
    \hat{d}_1 = \hat{b}_1e^{-i\phi_1}\cosh R
      + \hat{b}_2^\dag e^{i(\phi_2+2i\vartheta)}\sinh R \,,\\
    \hat{d}_2 = \hat{b}_2e^{-i\phi_1}\cosh R
      + \hat{b}_1^\dag e^{i(\phi_2+2i\vartheta)}\sinh R
  \end{gather}
\end{subequations}
where $R$, $\vartheta$ are, respectively, the squeeze factor and the squeeze angle of the NOPA~2. The corresponding output photon numbers are equal to
\begin{subequations}
  \begin{gather}
    \hat{d}_1^\dag\hat{d}_1 = \hat{b}_1^\dag\hat{b}_1\cosh^2R
      + [\hat{b}_1^\dag\hat{b}_2^\dag e^{2i(\phi_+ + \vartheta)}
        + \hat{b}_1\hat{b}_2 e^{-2i(\phi_+ + \vartheta)}]\cosh R\sinh R
      + \hat{b}_2\hat{b}_2^\dag\sinh^2R \,, \\
    \hat{d}_2^\dag\hat{d}_2 = \hat{b}_2^\dag\hat{b}_2\cosh^2R
      + [\hat{b}_1^\dag\hat{b}_2^\dag e^{2i(\phi_+ + \vartheta)}
        + \hat{b}_1\hat{b}_2 e^{-2i(\phi_+ + \vartheta)}]\cosh R\sinh R
      + \hat{b}_1\hat{b}_1^\dag\sinh^2 R \,.
  \end{gather}
\end{subequations}
It follows from these equations, that independently on the incident quantum state, this measurement procedure gives information only on the common phase $\phi_+$ and therefore has no advantages in comparison with the single-arm SU(1,1) interferometer considered in Sec.\,\ref{sec:SU11_1}.

In principle, sensitivity to the differential phase $\phi_-$ can be recovered by measuring the NOPA~2 outputs by means of the homodyne detectors. However, taking into account that the more simple SU(2)-type measurement schemes, considered in sections \ref{sec:homodyne2} and \ref{sec:direct2}, allow to asymptotically saturate the QCRB, such a sophisicated scheme with two phase references (NOPA pump and the homodyne detectors local oscilator beams) hardly has any sense. Therefore, we do not consider it here.

\section{Conclusion}\label{sec:conclusion}

We analyzed here the sensitivity of squeezed light enhanced interferometers, using the unified approach, based on the convenient uncertainty relation type form of QCRB, presented in Sec.\,\ref{sec:QCRB_gen}. We have considered both linear (SU(2)) and non-linear (SU(1,1)) interferometers, as well as of the hybrid SU(2)/SU(1,1) schemes. We have shown, that assuming the condition \eqref{big_alpha}, which is obligatory for all realistic high-precision phase measurements, the sensitivity of both linear (SU(2)) and non-linear (SU(1,1)) interferometers is described by the following simple uniform equations having the structure that first appeared in the work \cite{Caves1981}.

In the case of the single-arm (both SU(2) and SU(1,1)) interferometers,
\begin{equation}\label{conc_phi}
  \mean{(\delta\phi)^2} \ge \frac{e^{-2r}}{4\mathcal{N}} \,,
\end{equation}
where
\begin{equation}\label{calN}
  \mathcal{N} = N + o(N) \,,
\end{equation}
$N$ is the mean number of photons at the phase shifting object and $r$ is the corresponding squeeze factor. The function $o(N)$ is defined in the standard way: $o(N)/N\to0$ if $N\to\infty$.

In the case of the two-arm SU(2) interferometer, the sensitivity for the common and differential phase shifts depends on the squeeze factors $r_1$ and $r_2$ at the bright and the dark ports, respectively:
\begin{equation}\label{conc_phi_pm}
  \mean{(\delta\phi_+)^2} \ge \frac{e^{-2r_1}}{4\mathcal{N}} \,,\quad
  \mean{(\delta\phi_-)^2} \ge \frac{e^{-2r_2}}{4\mathcal{N}} \,,
\end{equation}
where $\mathcal{N}$ is still defined by Eq.\,\eqref{calN}, but with $N$ being the total photon number in the both arms.

Concerning the SU(1,1) options, in principle, the input part of the SU(2) interferometer (the input beapsplitter  + squeezer(s)) can be replaced by the SU(1,1) type one, that is by the single non-degenerate optical parametric amplifier with the squeeze factor $r$. In this case, the sensitivity still is defined by Eqs.\,\eqref{conc_phi_pm}, but with $-r_1 = r_2 = r$. At the same time, the use of the SU(1,1)-type output part in the two-arm interferometer hardly has any sense because it gives information on the common phase $\phi_+$ only; therefore, a single-arm interferometer would be sufficient for this case.

\acknowledgments

This work was supported by the Russian Foundation for Basic Research grant 19-29-11003.

\appendix

\section{Fisher information matrix}\label{app:Fisher}

Using Eqs.\,\eqref{QCRB_rho_a} - \eqref{N_commute}, Eq.\,\eqref{log_deriv0} can be presented as follows:
\begin{equation}\label{log_deriv1}
  i(\hat{\rho}\hat{N}_j - \hat{N}_j\hat{\rho})
  = \frac{1}{2}(\hat{\rho}\hat{L}_j + \hat{L}_j\hat{\rho}) \,.
\end{equation}
Let $\ket{\rho_l}$, $p_l$ be the eigenstates and eigenvalues of $\rho$:
\begin{equation}
  \hat{\rho} = \sum_l\ket{\rho_l}p_l\bra{\rho_l} \,.
\end{equation}
Using this representation, Eq.\,\eqref{log_deriv0} can be solved explicitly:
\begin{equation}
  \bra{\rho_l}\hat{L}_j\ket{\rho_m}
  = 2i\frac{p_l-p_m}{p_l+p_m}\bra{\rho_l}\hat{N}_j\ket{\rho_m} \,.
\end{equation}
Substitution of this solution into Eq.\,\eqref{bbA} gives:
\begin{multline}\label{bbA1}
  \mathbb{A}_{jk} = \frac{1}{2}\sum_{lm}p_l\Bigl(
      \bra{\rho_l}\hat{L}_j\ket{\rho_m}\bra{\rho_m}\hat{L}_k\ket{\rho_l}
      + \bra{\rho_l}\hat{L}_k\ket{\rho_m}\bra{\rho_m}\hat{L}_j\ket{\rho_l}
    \Bigr) \\
  = 2\sum_{lm}\frac{(p_l-p_m)^2}{p_l+p_m}
      \bra{\rho_l}\hat{N}_j\ket{\rho_m}\bra{\rho_m}\hat{N}_k\ket{\rho_l} \,.
\end{multline}
Suppose that $\hat{\rho}$ is a pure state:
\begin{equation}
  p_0 = 1 \,, \quad p_{l\ne0} = 0 \,.
\end{equation}
In this case, only the terms with $l=0$, $m\ne0$ and $l\ne0$, $m=0$ survive in Eq.\,\eqref{bbA1}:
\begin{multline}
  \mathbb{A}_{jk} = 2\biggl(
      \sum_{m\ne0}\bra{\rho_0}\hat{N}_j\ket{\rho_m}\bra{\rho_m}\hat{N}_k\ket{\rho_0}
      + \sum_{l\ne0}\bra{\rho_l}\hat{N}_j\ket{\rho_0}\bra{\rho_0}\hat{N}_k\ket{\rho_l}
    \biggr) \\
  = 4(\mean{\hat{N}_j\hat{N}_k} - \mean{\hat{N}_j}\mean{\hat{N}_k}
  = 4\mean{\delta\hat{N}_j\delta\hat{N}_k} \,.
\end{multline}

\section{Single-arm interferometer with SU(1,1) measurement}\label{app:SU_1_1}

Evolution of the optical field in the single-arm SU(1,1) interferometer shown in  Fig.\,\ref{fig:single}(a) is described by Eqs.\,(\ref{b_1}, \ref{d_SU11}). It follows from these equations that
\begin{equation}
  \hat{d} = B\alpha + C\hat{z} + S\hat{z}^\dag \,,
\end{equation}
where
\begin{subequations}
  \begin{gather}
    B = e^{-i\phi}\cosh R - e^{2i\theta+i\phi}\sinh R \,, \\
    C = e^{-i\phi}\cosh r\cosh R - e^{i\phi}\sinh r\sinh R \,, \\
    S = e^{2i\theta-i\phi}\sinh r\cosh R - e^{2i\theta+i\phi}\cosh r\sinh R \,.
  \end{gather}
\end{subequations}
Therefore, the mean value and variance of $\hat{n}=\hat{d}^\dag\hat{d}$ are equal to
\begin{subequations}\label{n_SU_1_1}
  \begin{gather}
    \mean{\hat{n}} = |B|^2\alpha^2 + |S|^2 \,, \\
    \mean{(\delta\hat{n})^2} = |B^*S + BC^*|^2\alpha^2 + 2|C|^2|S|^2 \,.
  \end{gather}
\end{subequations}
In the particular case of \eqref{big_R},
\begin{subequations}
  \begin{gather}
    B = \frac{e^R}{2}(e^{-i\phi} - e^{2i\theta+i\phi}) \,, \\
    C = \frac{e^R}{2}\sigma \,, \\
    S = -\frac{e^R}{2}e^{2i\theta}\sigma^*\,,
  \end{gather}
\end{subequations}
where
\begin{equation}
  \sigma = e^{-i\phi}\cosh r - e^{i\phi}\sinh r \,,
\end{equation}
which gives Eqs.\,\eqref{SU_1_1_stat}.

\section{Variances of the photon numbers in the arms of the SU(2) interferometer}\label{app:dN_pm_SU2}

It follows from Eqs.\,\eqref{a12_SU2} that
\begin{subequations}
  \begin{gather}
    \delta(\hat{a}_j^\dag\hat{a}_j)\ket{0} = \bigl[
        (\alpha_jG_j + \alpha_j^*g_j)\hat{z}_j^\dag + G_jg_j\hat{z}_j^\dag{}^2
      \bigr]\ket{0} \,, \\
    \delta(\hat{a}_1^\dag\hat{a}_2)\ket{0} = \bigl(
        \alpha_2G_1\hat{z}_1^\dag + \alpha_1^*g_2\hat{z}_2^\dag
        + G_1g_2\hat{z}_1^\dag\hat{z}_2^\dag
      \bigr)\ket{0} \,, \\
    \delta(\hat{a}_2^\dag\hat{a}_1)\ket{0} = \bigl(
        \alpha_1G_2\hat{z}_2^\dag + \alpha_2^*g_1\hat{z}_1^\dag
        + G_2g_1\hat{z}_1^\dag\hat{z}_2^\dag
      \bigr)\ket{0} \,,
  \end{gather}
\end{subequations}
and
\begin{subequations}
  \begin{gather}
    \delta\hat{N}_+\ket{0}
      = \sum_{j=1,2}\bigl(A_{jj}\hat{z}_j^\dag + G_jg_j\hat{z}_j^\dag{}^2\bigr)\ket{0} \,, \\
    \delta\hat{N}_-\ket{0} = \bigl[
        A_{21}\hat{z}_1^\dag + A_{12}\hat{z}_2^\dag
        + (G_1g_2 + G_2g_1)\hat{z}_1^\dag\hat{z}_2^\dag
      \bigr]\ket{0} \,,
  \end{gather}
\end{subequations}
where
\begin{gather}
  G_j = \cosh r_j \,, \quad g_j = e^{2i\theta_j}\sinh r_j \,, \\
  A_{jk} = \alpha_jG_k + \alpha_j^*g_k \,.
\end{gather}
Therefore,
\begin{subequations}\label{DN_pm_gen}
  \begin{gather}
    \mean{(\delta\hat{N}_+)^2} = \sum_{j=1,2}\bigl(|A_{jj}|^2 + 2G_j^2|g_j|^2\bigr) , \\
    \mean{(\delta\hat{N}_-)^2} = |A_{21}|^2 + |A_{12}|^2 + |G_1g_2 + G_2g_1|^2 \,, \\
    \mean{\delta\hat{N}_+\delta\hat{N}_-} = A_{11}^*A_{21} + A_{22}^*A_{12} \,.
  \end{gather}
\end{subequations}
With account for the conditions \eqref{alpha12}, Eqs.\,\eqref{DN_pm_gen} reduce to Eqs.\,\eqref{DN_pm_raw}.

\section{Measurement error in the double direct detection case}\label{app:direct2}

Eqs.\,\eqref{a12_SU2meas} give that
\begin{subequations}
  \begin{gather}
    \hat{n}_1\ket{0} = (
        \alpha^2 + \alpha\hat{z}_1^\dag e^{r_1} + \hat{z}_1^\dag{}^2\cosh r_1\sinh r_1
        + \sinh^2r_1
      )\ket{0} \,, \\
    \hat{n}_2\ket{0} = (\hat{z}_2^\dag{}^2\cosh r_2\sinh r_2 + \sinh^2r_2)\ket{0} \,, \\
    \hat{Y}\ket{0} = i[
        \alpha\hat{z}_2^\dag e^{-r_2} + \hat{z}_1^\dag\hat{z}_2^\dag\sinh(r_1 - r_2)
      ]\ket{0} \,,
  \end{gather}
\end{subequations}
Therefore,
\begin{subequations}
  \begin{gather}
    \mean{\hat{n}_1} = \alpha^2 + \sinh^2r_1 \,, \\
    \mean{\hat{n}_2} = \sinh^2r_2 \,, \\
    \mean{Y} = 0 \,,
  \end{gather}
\end{subequations}
\begin{subequations}
  \begin{gather}
    \mean{(\delta\hat{n}_1)^2} = \alpha^2e^{2r_1} + \frac{\sinh^22r_1}{2} \,,  \\
    \mean{(\delta\hat{n}_2)^2} = \frac{\sinh^22r_2}{2} \,, \\
    \mean{\delta\hat{n}_1\circ\delta\hat{n}_2} = 0 \,, \\
    \mean{(\delta\hat{Y})^2} = \alpha^2e^{-2r_2} + \sinh^2(r_1 - r_2) \,, \\
    \mean{\delta\hat{n}_1\circ\delta Y} = \mean{\delta\hat{n}_2\circ\delta Y} = 0 \,,
  \end{gather}
\end{subequations}
where ``$\circ$'' denotes the symmetrized product. Then, using Eqs.\,(\ref{n_pm_out}, \ref{n_pm}):
\begin{equation}
  \mean{\hat{n}_-} = \alpha^2 + \sinh^2r_1 - \sinh^2r_2 \,,
\end{equation}
\begin{subequations}
  \begin{gather}
    \mean{(\delta\hat{n}_+)^2} = \mean{(\delta\hat{n}_-)^2}
      = \mean{(\delta\hat{n}_1)^2} + \mean{(\delta\hat{n}_2)^2} \,, \\
    \mean{\delta\hat{n}_-\delta\hat{n}_+}
      = \mean{(\delta\hat{n}_1)^2} - \mean{(\delta\hat{n}_2)^2} \,, \\
    \mean{\delta\hat{n}_+\circ\delta\hat{Y}}
      = \mean{\delta\hat{n}_-\circ\delta\hat{Y}} = 0 \,,
  \end{gather}
\end{subequations}
and
\begin{subequations}
  \begin{gather}
    \mean{\hat{n}_-^{\rm out}} = \mean{n_-}\cos2\phi_- \,,  \\
    G = -2\mean{\hat{n}_-}\sin2\phi_- \,, \label{G_direct_2}
  \end{gather}
\end{subequations}
\begin{subequations}\label{n_pm_direct_stat}
  \begin{gather}
    \mean{(\delta\hat{n}_+^{\rm out})^2}
      = \mean{(\delta\hat{n}_1)^2} + \mean{(\delta\hat{n}_2)^2} \,, \\
    \mean{(\delta\hat{n}_-^{\rm out})^2}
      = [\mean{(\delta\hat{n}_1)^2} + \mean{(\delta\hat{n}_2)^2}]\cos^22\phi_-
      + \mean{(\delta\hat{Y})^2}\sin^22\phi_- \,,
    \\
    \mean{\delta\hat{n}_-^{\rm out}\delta\hat{n}_+^{\rm out}}
      = (\mean{(\delta\hat{n}_1)^2} - \mean{(\delta\hat{n}_2)^2})\cos2\phi_- \,.
  \end{gather}
\end{subequations}
Substitution of Eqs.\,(\ref{G_direct_2}, \ref{n_pm_direct_stat}) into Eq.\,\eqref{dphi_direct_raw} gives that
\begin{equation}\label{dphi_direct_gen}
  \mean{(\delta\phi_-)^2}
  = \frac{1}{4\mean{\hat{n}_-}^2}\biggl[
        \mean{(\delta\hat{Y}^2)}
        + \frac{4\mean{(\delta\hat{n}_1)^2}\mean{(\delta\hat{n}_2)^2}}
            {\mean{(\delta\hat{n}_1)^2} + \mean{(\delta\hat{n}_2)^2}}
            \cot^22\phi_-
      \biggr] .
\end{equation}


%

\end{document}